\documentstyle[epsf]{europhys} 

\def\stars{\bigskip\centerline{***}\medskip}

\newif\ifboo \boofalse

\newcommand{\text}[1]{\hbox{\rm #1}}

\begin{document} 
\euro{}{}{}{} 
\Date{} 
\shorttitle{A. Silva and S.Levit,
Controlled Dephasing of a Quantum Dot in the Kondo Regime. } 
\title{ Controlled Dephasing of a Quantum Dot in the Kondo Regime.}
\author{Alessandro Silva and Shimon Levit}   
\institute{Dept. of Condensed Matter Phys., The Weizmann Institute of Science, 
76100 Rehovot, Israel}

\rec{}{in final form } 
  
\pacs{
\Pacs{75}{20.Hr}{Kondo effect}   
\Pacs{73}{23.-b}{Mesoscopic systems}
\Pacs{03}{65.Yz}{Decoherence} 
}

\maketitle
\begin{abstract}
{In this work we analyze how coherent transport through a Quantum Dot (QD) in the
Kondo regime is affected by the weak capacitive interaction with a nearby biased Quantum 
Point Contact (QPC). We find that when the QD-QPC interaction is weak 
the width of the Kondo resonance is hardly affected by it. 
However, the spectral weight of the Kondo peak is 
reduced leading to an observable suppression of the conductance
and the elastic transmission probability through the QD.}  
\end{abstract}


\vspace{0.5cm}
  
Recently, some interesting developments in the study of Quantum Dots
(QD) have emerged, such as the 
theoretical prediction~\cite{Glazman} and recent 
observation~\cite{Goldhaber} of the Kondo effect in these systems.
Indeed, when a QD is on the average occupied by an odd number
of electrons and at temperatures below a characteristic scale
$T_k$, the Kondo temperature, the strong Coulomb interaction 
produces a many-body resonance at the Fermi level~\cite{Hewson}
which manifests itself in an enhancement of the conductance through the QD.
At temperatures 
$T \ll T_k$ the conductance $G$ reaches its unitary limit ($G \approx 2e^2/h$).
The realization of the Kondo effect in QD's permits 
the consideration of many aspects of Kondo
physics (e.g., the Kondo effect out of equilibrium~\cite{Meir2,Glazman2})
whose study is experimentally difficult in traditional realizations of 
the Kondo effect, such as magnetic impurities in a bulk metallic host.

Another interesting development has been the experimental demonstration
of the coherence of transport through a 
QD~\cite{Yacoby}, as well as the realization 
of \bf controlled dephasing \rm experiments~\cite{Bucks}.
In the latter, one studies the influence  of the interaction with a
controlled environment (the ``dephasor'') on coherent transport through a 
nanoscale system (the ``dephasee''). 
Recently, 
the situation where
the ``dephasee'' is a QD tuned at a Coulomb Blockade (CB) peak, 
while  the role of the ``dephasor'' is played
by a biased quantum point contact (QPC)
placed nearby  has been considered 
both experimentally~\cite{Bucks}
and theoretically~\cite{Yehoshua,Alessandro}. 
The resulting dephasing was detected experimentally
by inserting the QD in one arm
of an Aharonov-Bohm (AB) interferometer \cite{Bucks}.

When a QD tuned close to a CB resonance is interacting with a biased
QPC the only many-body effect on 
the dephasing are the restrictions imposed by the Pauli principle on the 
scattering of electrons in the two subsystems~\cite{Yehoshua,Alessandro}.
On the other hand, if a QD is tuned inside a ``Kondo valley'' at $T \ll T_K$,
the system consisting of the QD and the metallic leads attached to it 
is in a strongly correlated many-body state to start with.
In order to probe the nature of these correlations, it is  appealing
to study the qualitative differences between the response of 
a Kondo resonance and a  CB resonance to the interaction with an 
environment. 
In this Letter, we study how the coherent transport through a QD in the Kondo
regime is affected by the \bf capacitive \rm interaction with an
environment (a biased QPC).
We find that for weak QD-QPC interaction the width of the Kondo 
resonance is hardly affected by the environment. However,
the spectral weight of the resonance is reduced leading to 
an observable suppression of transport properties, such as the conductance. This
result has to be contrasted with the effect of the interaction with a biased QPC
on a QD at a CB peak, i.e. the suppression of the transport properties
through the inelastic broadening of the CB resonance~\cite{Yehoshua,Alessandro}. 

{\setlength{\unitlength}{1cm}
\begin{figure}
\vglue 0.5cm \epsfxsize=0.48\hsize
\hspace{4cm}
\begin{picture}(3,3)
      \epsffile{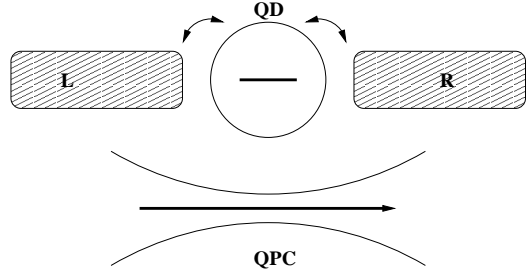}
   \end{picture}
\vspace{0cm} \caption{ Schematic picture of the system under consideration:
two metallic leads, left (L) and right (R), are connected by a quantum
dot (QD) through which electrons can tunnel. Electrons in a nearby 
biased quantum point contact (QPC)  interact capacitively 
with the electrons in the QD.} \label{Fig1}
\end{figure}}

The system under consideration is depicted schematically in Fig.(\ref{Fig1}).
Assuming for simplicity the QD coupled symmetrically to 
two metallic leads and described by the usual Anderson model,
its Hamiltonian is 
\begin{eqnarray}\label{H0}
&&\hat{H}_0=\hat{H}_{leads}+
\sum_{\sigma} \epsilon_{0} \; d^{\dagger}_{\sigma}
 d_{\sigma}
+\sum_{k,\sigma} W_0 \left[\;c^{\dagger}_{\sigma,k}(i)\;
d_{\sigma}+ h.c.\right]+ U\;n_{\downarrow}n{\uparrow},
\end{eqnarray}
where $\hat{H}_{leads}=\sum_{k,\sigma}\;
\epsilon_k\; c^{\dagger}_{\sigma,k}(i) 
c_{\sigma,k}(i)$ is the Hamiltonian of the leads,
the operators $c_{\sigma,k}(i)$ refer to states in the
leads ($i=L,R$), and 
$d_\sigma$ annihilates an electron with spin $\sigma$
in the QD. The local Coulomb interaction is represented
by the last term, where $n_{\sigma}=d^{\dagger}_{\sigma}d_{\sigma}$.
We will assume the density of states in the
leads $\rho$ to be constant with half width $D$.

Describing the QPC in terms of scattering states~\cite{Lesovik}, 
the total Hamiltonian takes the form $\hat{H}=\hat{H}_0+\hat{H}_{QPC}+\hat{V}$, where 
\begin{eqnarray}\label{HQPC} 
&& \hat{H}_{QPC}=\sum_{q,j,\alpha}\; 
\epsilon_q\;a^{\dagger}_{q\alpha}(j)a_{q\alpha}(j)~,\\
&& \hat{V}= \sum_{\sigma} n_{\sigma}\;\left[
 \sum \; V_{q,q^{\prime}}(j,j^{\prime})
a^{\dagger}_{q \alpha}(j)a_{q^{\prime}\alpha}(j^{\prime})\right]~.\label{Int}
\end{eqnarray}
Here the operators $a^{\dagger}_{q\alpha}(j)$ ($a_{q\alpha}(j)$)
create (annihilate) left-right going scattering states 
($j=L,R$) with spin $\alpha=\uparrow,\downarrow$.
The application of a bias shifts the chemical potentials of the right-left going
scattering states such that $\mu_R-\mu_L=eV$. 
In the following we take the matrix elements
$V_{q,q}(j,j)$ to be zero if $\epsilon_{q}<\mu_L$, and equal to a constant otherwise
($V_{q,q^{\prime}}(j,j^{\prime})=V$) . This corresponds to the inclusion 
of the equilibrium Hartree shift in the definition of $\epsilon_0$.

In order to characterize the influence of the QPC on transport through the
QD it is convenient to calculate the elastic transmission amplitude 
$t\rm^{el}_d(\omega)$  
and the total transmission probability $\cal{T}\rm_{d}(\omega)$ 
through the QD at zero temperature. 
In terms of the QD Green's function
$G_{d}^{r}(t)=-i\; \theta(t)\; \langle \{d_{\sigma}(t),d^{\dagger}_{\sigma}(0) \}\rangle$, 
these quantities are given by
\begin{eqnarray}\label{amplitude}
&&  t\rm^{el}_d(\omega)=-i \Gamma_0\; G^{r}_{d}(\omega),\\
&&\cal{T}\rm_{d}(\omega)= (\Gamma_0/2)\; A_d(\omega),
\label{probability}
\end{eqnarray} 
where $\Gamma_0=\Gamma_L+\Gamma_R=2\pi \mid W_0 \mid^2 \rho$, and 
$A_d(\omega)=-2\; {\rm Im}[G_d^{r}(\omega)]$ is the dot's spectral density.
At zero temperature and in the 
absence of interaction with the QPC (the unitary limit), one has 
$\cal{T}\rm_{d}(\omega)=\mid t^{el}_d(\omega)\mid^2$. 
At $T=0$ the linear response conductance through the QD is proportional to  
total transmission probability at the Fermi level $G=(2e^2/h)\cal{T}\rm_{d}(0)$,
where we set $\epsilon_F=0$~\cite{Meir}. 
Similarly, if the QD is inserted in one arm of an AB 
interferometer, the amplitude $\cal{A}$ of the flux-sensitive oscillatory component of the
conductance is set by the absolute value of the elastic transmission amplitude
at the Fermi level, $\cal{A}\rm \propto \mid t^{el}_d(0)\mid $~\cite{Oreg}. 
In principle, one can also study experimentally  $\cal{T}\rm_{d}(\omega)$ for $\omega>0$ 
by measuring the conductance through the
QD for electrons injected in the leads above the Fermi level by means 
of an energy filter.  

We are interested in the response of
the QD in the Kondo regime (i.e. $\epsilon_0 \ll 0$, $T \ll T_k$), 
when the Kondo resonance is weakly perturbed by the 
interactions with the QPC. In order to 
incorporate at lowest order the main qualitative features 
of the Kondo effect 
we choose to perform our calculation in the framework
of Slave-Boson Mean Field Theory \cite{Coleman}. 

We start by considering the  
Anderson Hamiltonian, Eq.(\ref{H0}), in the $U \rightarrow \infty$ limit. 
In this case, it is possible to show \cite{Coleman}
that the $(N=2)$ - fold Anderson model can be represented in terms of a new set of 
operators, a spin less slave-boson $b$ and two pseudo-fermions $f_{\sigma}$,
satisfying the constraint 
$Q=b^{\dagger}b+ \sum_{\sigma} f^{\dagger}_{\sigma}f_{\sigma}=N/2$ 
($\sigma=1,..,N$). 
Introducing a Lagrange multiplier $\lambda$ the Slave-Boson representation of the
Hamiltonian is $\hat{H}=\hat{H}_{SB}+\hat{H}_{QPC}+\hat{V}_{SB}$, where
\begin{eqnarray}\label{HSB}
&&\hat{H}_{SB}=\hat{H}_{leads}+ \sum \epsilon_{0} \; f^{\dagger}_{\sigma}
f_{\sigma} +\sum_{k,\sigma} W_0 (\;c^{\dagger}_{\sigma,k}(i)\;
b^{\dagger} f_{\sigma}+ h.c.) + \lambda (Q-N/2),\\ 
&&\hat{V}_{SB}=\sum_{\sigma}f^{\dagger}_{\sigma} f_{\sigma} \;\left[
 \sum \; V_{q,q^{\prime}}(j,j^{\prime})
a^{\dagger}_{q \alpha}(j)a_{q^{\prime}\alpha}(j^{\prime})\right]~,
\end{eqnarray}
and $\hat{H}_{QPC}$ is still given by Eq.(\ref{HQPC}).
  
In terms of the slave-boson representation of the 
Hamiltonian the low temperature behavior of the 
Kondo problem is 
recovered by means of a simple Mean Field
Theory (MFT). This amounts to the treatment of the bosonic field as
static c-number by means of the substitution $b \rightarrow 
\langle b \rangle=\tilde{b}$. At $T=0$ this approximation describes
properly the low energy spin-fluctuations and is exact in
the $N \rightarrow +\infty$ limit \cite{Coleman}.
In particular, the MF Hamiltonian takes the form $\hat{H}_{MF}+\hat{H}_{QPC}+\hat{V}_{SB}$, 
where the first term, up to a constant, is given by
\begin{eqnarray}\label{HMF}
&&\hat{H}_{MF}= \hat{H}_{leads}+ \sum \epsilon \; f^{\dagger}_{\sigma}
 f_{\sigma}+
+\sum \left[W^{*} \;c^{\dagger}_{\sigma,k}(i)\;
f_{\sigma}+ h.c.\right]. 
\end{eqnarray}
The MF Hamiltonian is easily recognized to describe a resonant level 
of renormalized energy $\epsilon=\epsilon_0+\lambda$ interacting with the QPC, 
and coupled to the two leads by a mean field dependent coupling 
$W=W_0 \tilde{b}$. 

The mean fields $\lambda$ and $\tilde{b}$ are determined 
self-consistently by means of two MF equations
$\lambda\tilde{b}+\sum_{k\;i \sigma} W_0 \langle c^{\dagger}_{k,\sigma}(i) 
f_{\sigma}\rangle=0$ and $\mid \tilde{b} \mid^2+\sum_{\sigma}\langle 
f^{\dagger}_{\sigma}f_{\sigma}\rangle=\frac{N}{2}$, 
where the first can be obtained from the equations of motion of $b(t)$ and 
the second enforces the constraint at the MF level ($\langle Q \rangle=N/2$).
It is now possible, by means of standard techniques~\cite{Meir}, 
to integrate out the leads and express 
the MF equations
only in terms of 
the Fourier transforms of the pseudo-fermion Green's functions 
$F^r_{\sigma}(t)=-i\theta(t)\; 
\langle \{f_{\sigma}(t),f^{\dagger}_{\sigma}(0) \}\rangle$ and $F^{<}_{\sigma}(t)=i 
\langle f^{\dagger}_{\sigma}(0)f_{\sigma}(t) \rangle$.
After some straightforward manipulations one obtains
\begin{eqnarray}\label{selfcon1}
(\epsilon-\epsilon_0)+&& \Gamma_0 \sum_{\sigma}\int \frac{d\omega}{2\pi} 
\left[ 2F^{r}_{\sigma}(\omega) f(\omega)+F^{<}_{\sigma}(\omega) \right]=0,\\
&&\frac{\Gamma}{\Gamma_0}+\sum_{\sigma}\int\;\frac{d\omega}{2\pi i}
F^{<}_{\sigma}(\omega)=\frac{N}{2}\;,\label{selfcon2} 
\end{eqnarray}
where $\Gamma=\Gamma_0 \mid \tilde{b} \mid^2$, and $f(\omega)$ is the Fermi function. 
In particular, from the knowledge of $\Gamma/\Gamma_0 
=\mid \tilde{b} \mid^2$ and of the pseudo-fermion Green's function
$F^{r}_{\sigma}(\omega)$, one can find the MF expression for the 
QD Green's function 
as $G^{r}_d(\omega)=\mid \tilde{b} 
\mid^2 F^r_{\sigma}(\omega)$ and finally obtain the transmission amplitudes
through Eq.(\ref{amplitude})-(\ref{probability}).

For later purposes, it is convenient to summarize a few results 
in the \it absence of interaction \rm with the QPC. In this case, 
the MF Hamiltonian Eq.(\ref{HMF})
is quadratic and the Green's functions $F^{r}_{\sigma},F^{<}_{\sigma}$ can be easily 
calculated. The MF equations 
take the simple form~\cite{Coleman}
\begin{eqnarray}\label{nonintMF1}
(\epsilon-\epsilon^*)+&&\frac{N \Gamma_0}{\pi}\ln\left[\frac{
\sqrt{(\epsilon)^2+(\Gamma)^2}}{\Gamma_0}\right]=0,\\
&&\frac{\Gamma}{\Gamma_0}+\frac{N}{\pi}\arctan\left[\frac{\Gamma}
{\epsilon}\right]=\frac{N}{2},\label{nonintMF2}
\end{eqnarray}
where  $\epsilon^*=\epsilon_0+N\;\Gamma_0/\pi\;\ln\left[D/\Gamma_0\right]$
is the Haldane invariant level~\cite{Haldane}. In the Kondo regime 
($\epsilon^*\ll-\Gamma_0$) one finds that the solution describes a 
resonance close to the Fermi level with
\begin{eqnarray}\label{solutionNI}
\Gamma=T_k=\Gamma_0\;\exp\left[\frac{\pi\epsilon^*}{N \Gamma_0}\right]
\;\;\;\;\;\;\;0<\epsilon \ll T_k.
\end{eqnarray}

In the \it presence of interaction \rm with the QPC,  
in order to close self-consistently the MF equations one has 
to calculate the interacting pseudo-fermion Green's functions. 
This task can 
be achieved by a standard real-time perturbation expansion in the interaction $\hat{V}$
between the QD and the QPC. In terms of the proper self-energies the pseudo-fermion
Green's functions can be written in the form
\begin{eqnarray}\label{PTresult1}
F^{r}_{\sigma}(\omega)&=&\frac{1}{\omega-\epsilon+i \Gamma-\Sigma^{r}},\\
F^{<}_{\sigma}(\omega)&=&i\;A(\omega)\;\left[f(\omega)+\delta f(\omega) \right],
\label{PTresult2}
\end{eqnarray}
where $f(\omega)$ is the Fermi function, $A(\omega)=-2 {\rm Im}[F^r(\omega)]$
is the interacting spectral density, and 
\begin{equation}\label{distribution}
\delta f(\omega)=\frac{(1-f(\omega))\Sigma^{<}(\omega)+f(\omega)\Sigma^{>}(\omega)}
{2\;i\;(\Gamma+\Gamma_d(\omega))},
\end{equation}
the change in the distribution function,
where $\Gamma_d(\omega)=-{\rm Im}[\Sigma^r(\omega)]$.

For weak interaction,  it is enough 
to evaluate the self-energies up to second order in the
QD-QPC interaction~\cite{Alessandro}. 
For the lesser and greater self-energies one easily obtains
\begin{eqnarray}\label{lesser}   
\Sigma^{<}(\omega)&=&2i\int \frac{d\omega^{\prime}}{2\pi}A_0(\omega^{\prime})
f(\omega^{\prime}) \rho_0(\omega^{\prime}-\omega),\\
\Sigma^{>}(\omega)&=&-2i\int \frac{d\omega^{\prime}}{2\pi}A_0(-\omega^{\prime})
f(\omega^{\prime}) \rho_0(\omega^{\prime}+\omega).\label{greater}
\end{eqnarray}
Here the spectral density $\rho_0$ describes particle-hole excitations in the QPC
and is given by  
\begin{eqnarray}\label{rho}
\rho_0 (\omega)&=& v^2 \sum_{j,j^{\prime}} \; 
\int d\omega_1 d\omega_2\; \delta(\omega_1-\omega_2-\omega)
f^{(j)}(\omega_2)\left(1-f^{(j^{\prime})}(\omega_1)\right) ~,
\end{eqnarray}
where we introduced the dimensionless couplings 
$v=L \mid V\rm \mid/(2 \pi v_F)$, and the the thermal
factor $f^{(j)}(\omega)=f(\omega-\mu_j)$ distinguishes left- and right-going scattering
states by their chemical potential difference.

The retarded self-energy can be expressed in terms of $\Sigma^{<},\Sigma^{>}$ as
\begin{equation}\label{retarded}
\Sigma^{r}(\omega)=\Delta+i\;\int \frac{d\omega^{\prime}}{2\pi} \frac{\Sigma^{>}
(\omega^{\prime})-\Sigma^{<}(\omega^{\prime})}{\omega-\omega^{\prime}+i\delta},
\end{equation}
where $\Delta=2 v eV$ is the first order shift. In particular 
$\Gamma_d(\omega)=(\Sigma^{>}(\omega)-\Sigma^{<}(\omega))/2$.

For weak interaction ($v \ll 1$), the solution of Eq.(\ref{selfcon1}),
Eq.(\ref{selfcon2}) will give  a small correction to the noninteracting solutions 
. Therefore, in the Kondo regime we expect the peak of the pseudo-fermion spectral 
density to 
be close to the Fermi level (see, Eq.(\ref{solutionNI})). Taking advantage of this,
one can find a closed expression for the self-energies in the relevant frequency range 
($\mid \omega \mid < \Gamma$), and therefore for the Green's functions of 
Eq.(\ref{PTresult1}) and Eq.(\ref{PTresult2}). 
Indeed, provided $0<\epsilon <\Gamma$ and $eV \gg \Gamma$,
it is easy to prove~\cite{Alessandro} that the function $\Gamma_d$ is a smooth 
function of $\omega$ on a scale $\Gamma$ around the Fermi level and 
can be approximated as 
\begin{equation}\label{Gammad}
\Gamma_d \simeq 2 \pi v^2 \mid eV \mid.
\end{equation}
In particular $\Gamma_d$ is assumed to be a small correction
to $\Gamma$ ($\Gamma_d \ll \Gamma$).
Similarly for the real part of $\Sigma^r$, one can retain only the first order
shift $\Delta$, since the second order contribution is approximately 
zero for $0 < \epsilon \ll \Gamma$. Therefore, we obtain the 
simple expression
\begin{equation}\label{retarderfinal}
F^{r}_{\sigma}(\omega)=\frac{1}{\omega-\epsilon-\Delta+
i (\Gamma+\Gamma_d)}.
\end{equation}

Another simplification of the problem comes from the fact that for 
$0 < \epsilon \ll \Gamma $ one obtains that 
$\Sigma^{>}(-\omega)\simeq -\Sigma^{<}(\omega)$. In turn, this implies that the correction to
the distribution function $\delta f(\omega)$ is an odd function of
$\omega$. Therefore, up to corrections $O(v^3)$ we have that 
\begin{equation}\label{integral}
\int \frac{d\omega}{2\pi}\;A(\omega)\delta f(\omega) \simeq 0.
\end{equation}  

Using Eq.(\ref{retarderfinal}) to obtain the interacting spectral density 
$A(\omega)$ and making use of Eq.(\ref{integral}) one can close the 
MF equations. In terms of the renormalized level $\epsilon_1=\epsilon+\Delta$
and of the renormalized total width $\Gamma_1=\Gamma+\Gamma_d$ 
(see Eq.(\ref{retarderfinal})) one obtains
\begin{eqnarray}\label{finalMF1}
(\epsilon_1-\epsilon^*)+&&\frac{N \Gamma_0}{\pi}\ln\left[\frac{
\sqrt{(\epsilon_1)^2+(\Gamma_1)^2}}{\Gamma_0}\right]=0,\\
&&\frac{\Gamma_1}{\Gamma_0}+\frac{N}{\pi}\arctan\left[\frac{\Gamma_1}
{\epsilon_1}\right]=\frac{N}{2}+\frac{\Gamma_d}{\Gamma_0},\label{finalMF2}
\end{eqnarray}
where we redefined the Haldane invariant level $\epsilon^*=\epsilon_0+\Delta
+N\;\Gamma_0/\pi\;\ln\left[D/\Gamma_0\right]$ in order to take into account the 
shift induced by the presence of the QPC. 
 
Let us first observe that without $\Gamma_d/\Gamma_0$ in the right hand side of 
Eq.(\ref{finalMF2}), the two equations have exactly the same form
as the noninteracting MF equations (Eq.(\ref{nonintMF1}),(\ref{nonintMF2})),
with the solution given by Eq.(\ref{solutionNI}), $\Gamma_1=T_k$ and
$0<\epsilon_1\ll T_k$. The corrections due to $\Gamma_d$ can be found 
writing $\Gamma_1=T_k+\delta\Gamma$
and $\epsilon_1=\epsilon+\delta\epsilon$, linearizing in $\delta\Gamma$, $\delta\epsilon$,
and using $\Gamma_d/\Gamma_0< T_k/\Gamma_0 \ll 1$.
One readily obtains 
\begin{eqnarray}\label{deltaGamma} 
\delta\Gamma&\approx& 2 (\pi/N)^2 (T_k/\Gamma_0)^2 \Gamma_d \ll \Gamma_d ,\\
\delta\epsilon&\approx& (\pi/N) (\Gamma_d/\Gamma_0) T_k \ll T_k .
\end{eqnarray}\label{deltaepsilon}

Using these results the MF expression for the interacting real-electron Green's function
in the Kondo regime can be written as
\begin{eqnarray}\label{Greenresult}
G^{r}_d(\omega)&=& \mid \tilde{b} \mid^2 \left[\frac{1}{\omega-\epsilon_1+
i\Gamma_1}\right]
\simeq \frac{T_k-\Gamma_d}{\Gamma_0}
\left[\frac{1}{\omega+i(T_k+\delta \Gamma)}\right].
\end{eqnarray}
Let us make a few comments about this result. First of all notice that
the result has the structure $G^{r}_d=Z\;G^{r}_{qp}$, where $Z=(T_k-\Gamma_d)/\Gamma_0$
can be interpreted as a quasi-particle residue and $G^{r}_{qp}$ as the quasi-particle
Green's function. As anticipated in the introduction, the interaction induced broadening 
of the quasi-particle Green's function is anomalously small ($\delta\Gamma \ll \Gamma_d$)
and the main effect is the suppression of the residue by a factor $(1-\Gamma_d/T_k)$,
meaning that the main effect of the interaction is a transfer of spectral density out of the
Kondo peak.

The width of the Kondo resonance and the spectral weight are related
to two different physical quantities, respectively
the low energy quasiparticle spectrum and the overlap
between quasiparticle states and actual electrons
(see e.g. Ref.\cite{Hewson}). The result above indicates that
the main effect of the interaction with the QPC is a suppression of the latter,
with a consequent suppression of single particle properties 
(e.g. the conductance). However, the quasiparticle 
spectrum, and thus thermodynamics 
is hardly affected by the environment. One could 
interpret this as a consequence of 
the fact that the Kondo resonance 
is a result of fluctuations of the local spin
degrees of freedom in the QD. Since $\Gamma_d \ll T_k$ and the QD interacts with the 
QPC in the charge channel (capacitive density-density interaction) the local spin
dynamics is hardly affected by it.

The result for the total transmission probability and elastic transmission amplitude
can be readily obtained by substituting Eq.(\ref{Greenresult}) in 
Eq.(\ref{amplitude}) and (\ref{probability}). One obtains the final result
\begin{eqnarray}\label{interacting}
\cal{T}\rm_{d}(\omega)&\simeq&\left(1-\frac{\Gamma_d}{T_k}\right)\left[\frac{T_k^2}
{\omega^2+T_k^2}\right],\\
t^{el}_{d}(\omega)&\simeq&-i\left(1-\frac{\Gamma_d}{T_k}\right)\left[\frac{T_k}
{\omega+iT_k}\right],\label{interacting1}
\end{eqnarray}
where $\delta\Gamma \ll T_K$ has been neglected.
Using  Eq.(\ref{interacting})-(\ref{interacting1}) to  calculate the 
suppression of the linear response conductance and AB oscillations one obtains 
$\cal{G}\rm/\cal{G}\rm_0\simeq\cal{A}\rm/\cal{A}\rm_0\simeq(1-\Gamma_d/T_k)$,
where $\cal{G}\rm_0$ and $\cal{A}\rm_0$ are the values of both observables
at zero bias (in the QPC).

In order to gain some insight into these results, let us compare them
with the solutions of the same problem when the QD is tuned  \bf at a Coulomb Blockade
resonance \rm~\cite{Yehoshua,Alessandro}. In the absence of 
the interaction with the QPC and at $T=0$, the transmission through a
CB resonance is elastic and one has 
$\cal{T}\rm_{d}(\omega)=\mid t^{el}_{d}(\omega)\mid^2=\Gamma_0^2/(\omega^2
+\Gamma_0^2)$. If the interaction with the QPC is switched one obtains instead
\begin{eqnarray}\label{ninteracting}
\cal{T}\rm_{d}(\omega)&\simeq&\left[\frac{\Gamma_0^2}
{\omega^2+(\Gamma_0+\Gamma_d)^2}\right],\\
t^{el}_{d}(\omega)&\simeq&-i\left[\frac{\Gamma_0}
{\omega+i(\Gamma_0+\Gamma_d)}\right],\label{ninteracting1}
\end{eqnarray}
where $\Gamma_d$ is given by Eq.(\ref{Gammad}) for $eV \gg \Gamma_0$.
These expression imply a suppression of the linear response conductance and 
AB oscillations through the formula
$\cal{G}\rm/\cal{G}\rm_0\simeq\cal{A}\rm/\cal{A}\rm_0\simeq(1-\Gamma_d/\Gamma_0)$,
where $\Gamma_0$ is the width of the CB resonance.

Though this result 
is similar to the corresponding
result in the Kondo regime 
(with the substitution
$T_K \rightarrow \Gamma_0$), from the comparison of the corresponding expressions
of the total transmission probability in the CB and Kondo case (Eq.(\ref{ninteracting})
and (\ref{interacting}) respectively) as well as of the elastic transmission amplitude
(Eq.(\ref{ninteracting1}) and (\ref{interacting1}) respectively) one can infer 
that the mechanism responsible for the suppression of $G$ and $\cal{A}\rm$
in the two cases is different.
Indeed, while in the Kondo regime the effect of the interaction with the environment
is a transfer of spectral weight without any additional broadening,
in the CB case the main effect is just a broadening due to inelastic events.
This difference can not be revealed by means of linear response measurements. 

In conclusion, in this Letter we discussed how transport through a 
QD in the Kondo regime is affected by interactions with a biased QPC
by calculating the total transmission probability and the elastic transmission amplitude
through the QD. We have shown that in the weak dephasing regime 
($\Gamma_d \ll T_k$ but $eV>T_k$) the main effect of the interactions
is a suppression of the spectral weight of the Kondo resonance, while
the interaction-induced broadening of the quasiparticle pole is anomalously 
small. By means of a direct comparison we discussed the qualitative difference
between a CB resonance and a Kondo resonance as far as the response to 
the interaction with an environment is concerned.  

\stars
 
We wish to express our thanks to Y.Levinson, V. Belinicher, K.Kikoin 
and Y.Oreg for many useful discussions. 
We also thank M. Heiblum, D. Sprinzak, Y. Ji and M. Schechter 
for discussions. This work is supported by DIP Grant No. DIP-C7.1.

\vspace*{-4mm} 
\newcommand{\noopsort}[1]{} \newcommand{\printfirst}[2]{#1} 
  \newcommand{\singleletter}[1]{#1} \newcommand{\switchargs}[2]{#2#1}


\begin{thebibliography}{10} 
\bibitem{Glazman} L. I. Glazman and M. E. Raikh, JETP Lett. \bf 47\rm, 452 (1988);
T. K. Ng and P. A. Lee, Phys. Rev. Lett \bf 61\rm, 1768 (1988). 
\bibitem{Goldhaber} D. Goldhaber-Gordon, H. Shtrikman, D. Mahalu et al. ,
Nature \bf 391\rm, 156 (1998).
\bibitem{Hewson} A. C. Hewson, The Kondo Problem to Heavy Fermions  
(Cambridge University Press, Cambridge, 1993).
\bibitem{Meir2} N. S. Wingreen and Y.  Meir, Phys. Rev. B \bf 49\rm, 11040 (1993). 
\bibitem{Glazman2}A. Kaminski, Yu. V. Nazarov, and L. I. Glazman,  
Phys. Rev. Lett. \bf 83\rm, 384 (1999); A. Kaminski, Yu. V. Nazarov, and L.I. Glazman,
Phys. Rev. B \bf 62\rm, 8154 (2000); J. M. Elzerman et al., 
J. Low Temp. Phys \bf 118\rm, 375 (2000).
\bibitem{Yacoby} A. Yacobi, M. Heiblum, D. Mahalu, and H. Shtrikman, Phys. Rev. 
Lett. \bf74\rm, 4047 (1995);R. Schuster, E. Bucks, M. Heiblum, D. Mahalu, 
V. Umansky, and H Shtrikman, Nature (London) \bf 385\rm, 417 (1997).
\bibitem{Bucks} E. Buks, R. Schuster, M. Heiblum, D. Mahalu, 
V. Umansky, Nature \bf 391\rm, 871 (1998); D. Sprinzak, Bucks E., M. Heiblum, H. Shtrikman, 
Phys. Rev. Lett. \bf 84\rm, 5820 (2000).
\bibitem{Yehoshua} Y. Levinson, Europhys. Lett  \bf 39\rm, 299 (1997);
 I. L. Aleiner, N. S. Wingreen, Y. Meir, Phys. Rev.
Lett. \bf 79\rm, 3740 (1997); Y. Levinson, Phys. Rev. B \bf 61\rm, 4748 (2000); 
M. Buttiker, A. B. Martin,Phys. Rev. B \bf 61\rm, 2737 (2000).
\bibitem{Alessandro} A. Silva and S. Levit, Phys. Rev. B \bf 63\rm, 201309(R) (2001).
\bibitem{Lesovik} G. B. Lesovik, JETP Lett. \bf 49\rm, 591 (1989).
\bibitem{Meir} Y. Meir, N. S. Wingreen, Phys. Rev. Lett. \bf 68\rm, 2512 (1992).
\bibitem{Oreg} U. Gerland, J. von Delft, T. A. Costi, Y. Oreg, Phys. Rev. Lett. 
\bf 84\rm, 3710 (2000).
\bibitem{Coleman} N. Read and D. M. News, J. Phys. C \bf 16\rm, L1055 (1983); P. Coleman, 
Phys. Rev. B \bf 29\rm, 3035 (1984); P. Coleman, Phys. Rev. B \bf 35\rm, 5072 (1987).
\bibitem{Haldane} F. D. M. Haldane, Phys. Rev. Lett \bf 40\rm, 416 (1978). 
\end{thebibliography}
\end{document}